# Mirrorless focusing of XUV high-order harmonics

L. Quintard[1], V. Strelkov[2], J. Vabek[1], O. Hort[1, †], A. Dubrouil[1, ‡], D. Descamps[1], F. Burgy[1], C. Péjot[1], E. Mével[1], F. Catoire[1], and E. Constant[1, 3, *].

[1]*Université de Bordeaux, CNRS, CEA, Centre Laser Intenses et Applications (CELIA), 43 rue P. Noailles, 33400 Talence, France*

[2]*A M Prokhorov General Physics Institute of Russian Academy of Sciences, 38, Vavilova Street, Moscow 119991, Russia.*
Moscow Institute of Physics and Technology (State university), 141700 Dolgoprudny, Moscow Region, Russia.

[3]*Université de Lyon, Université Claude Bernard Lyon 1, CNRS, Institut Lumière Matière (ILM), 69622 Villeurbanne, France.*

[†] *present address: ELI Beamlines Project, Institute of physics, Czech Academy of Sciences, Na Slovance 1999/2, 182 21 Praha 8, Czech Republic*

[‡] *present address : Femtoeasy, Femto Easy SAS, Batiment Sonora, parc scientifique et technologique Laseris 1 , 33114 Le Barp, France*

\* Corresponding author: eric.constant@univ-lyon1.fr

**Abstract:** By experimentally studying high-order harmonic beams generated in gases, we show how the spatial characteristics of these ultrashort XUV beams can be finely controlled under standard generation conditions. For the first time, we demonstrate that these XUV beams can be emitted as converging beams and get thereby focused after generation. We study this mirrorless focusing using a spatially chirped beam that acts as a spatially localized probe located inside the harmonic generation medium.  We analyze the XUV beam evolution with an analytical model providing the beam characteristics and obtain very good agreement with experimental measurements. The XUV foci sizes and positions vary strongly with the harmonic order and the XUV waist can be located at arbitrarily large distances from the generating medium. We discuss how intense XUV fields can be obtained with mirrorless focusing and how such order-dependent XUV beam characteristics are compatible with broadband XUV irradiation and attosecond science.

# INTRODUCTION

It is now possible to generate XUV attosecond pulses via high-order harmonic generation (HHG) in gases with high photon flux (1 - 3). These XUV beams exhibit good spatial coherence but their spatial properties (4 - 11) appear now as more complex than previously considered and dedicated experiments are developed to understand and control the harmonic beam spatial properties (8 – 12). Moreover, applications involving nonlinear processes (13 – 16) induced by high-order harmonics require high XUV intensities and good focusability of the XUV beams that remain difficult to achieve. Improving the understanding and control of the spatial properties of the high-order harmonic beams is essential.

We study here the impact of the macroscopic conditions of HHG in gases on the XUV beams wavefront and spatial profiles. Experimentally, HHG is performed with a wavefront corrected fundamental beam and we observe a strong evolution of the XUV beam properties with harmonic order and generating conditions. A Gaussian model developed to infer the beam properties shows that both the divergence and beam waist position can be controlled. We demonstrate how and why the longitudinal position of the HHG gas medium affects dramatically the XUV beam.

In contrast to current understanding, we show that harmonic beams can converge after the generating medium and are then focused without any XUV optics. This XUV mirrorless focusing is further studied by performing HHG with a spatially chirped fundamental beam that acts as a probe localized directly inside the generation medium. These results demonstrate that converging or diverging XUV beams can be obtained by XUV wavefront control.

# RESULTS

**Characterization of the far field XUV beam size evolution.**

To perform HHG, we used a 40 fs FWHM, high energy fundamental pulse centered at 800 nm loosely focused in a neon jet (200 µm nozzle diameter) mobile in the longitudinal direction over $z_{jet}$=+-75 mm (z = 0 being the IR focus position). The IR beam is spatially filtered and a deformable mirror (ISP System) provides a $\lambda$/100 wavefront correction ensuring good wavefront control of the emitted XUV light. We analyze the spatial profile of each harmonic in the far field with a flat field spectrometer (see details in the method section).

Our experimental observations confirmed that the harmonic beam size in the far field evolves strongly with the jet position (4). Harmonics arising from the short quantum path (17) emission were predominant for all z positions. Those emitted via the long quantum path were very divergent and are not considered in this paper.

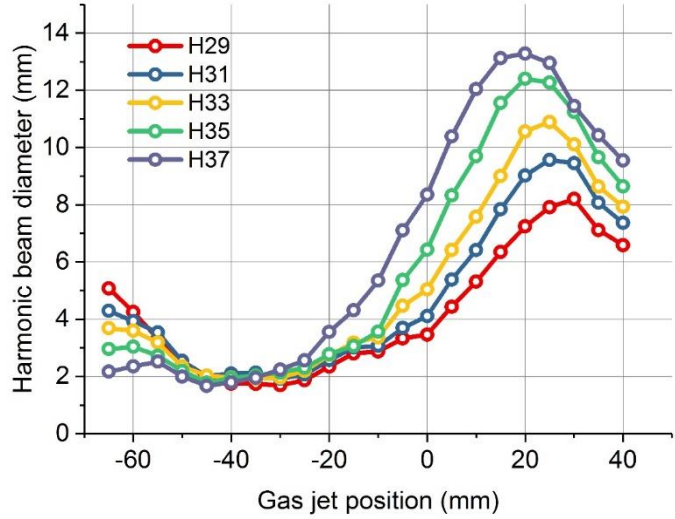

Figure 1: Measurement of the XUV beam size (FWHM) in the far field (i. e. 2.9 m after the IR focus) as a function of the longitudinal gas jet position with respect to the IR focus position. Harmonics are generated in a neon jet with a maximum intensity at focus of 5.6x10$^{14}$ W/cm$^2$.

The beam size (Fig. 1) of plateau harmonics evolves regularly with the jet position and harmonic order. It exhibits a shallow minimum between $z_{jet}$ = -30 mm (H29) and -40 mm (H35) followed by a maximum around $z_{jet}$ = 20 - 25 mm and a decrease afterward. We observed that the positions of the maxima and minima change with harmonic order. The maximum beam size (8 to 13 mm) increases with the harmonic order up to the order 37 and decreases afterward (see supplementary). The minimum beam size does not change significantly with the harmonic order but the position of the minima change.

Measurements showed also that the XUV photon flux was preserved between the beam size minima and maxima (see supplementary). The far field XUV beam shape can therefore be controlled without compromising on the XUV flux.

Despite the simplicity of this experiment, the physics underlying the beam evolution is poorly understood and phase matching, ionization or beam reshaping are often invoked to explain this evolution. To capture this physics and predict the harmonic beam characteristics, we developed a model including collective effects but neglecting longitudinal phase matching and beam reshaping.

**Simulations of the harmonic beam properties in the generating medium.**

The model that we developed is known to provide good qualitative results for HHG in a thin gas medium (18 – 23). The model dipole, $d_q$, of the harmonic with order q is given by:

$$d_q(r,z) = I_{ir}^{q_{eff}/2}(z,r) e^{i(q\varphi_{ir}(z,r) - \alpha_q I_{ir}(z,r))} \qquad (1)$$

Where $I_{ir}(z,r)$, is the intensity of the Gaussian fundamental beam at radial position *r* in the generation medium located at position z and $q_{eff}$ is an effective order of non-linearity. The phase, $\varphi_q$, of the harmonic dipole is *q* times the phase of the fundamental beam, $\varphi_{ir}$, plus the atomic contribution approximated by $\varphi_{atom} = -\alpha_q I_{ir}(z, r)$ (17).

This dipole provides the XUV field generated by an atom. The macroscopic response is obtained by considering an infinitely thin medium and by summing the contributions of many atoms located in the generation plane at various radial position. In this approach, we neglect propagation in the medium as justified with a thin medium and loosely focused laser (see methods). When the fundamental beam is Gaussian in the emission plane, with size $W_{ir}(z)$, the spatial profile of the XUV beam is also Gaussian in this plane with size:

$$W_{XUV}(z) = \frac{W_{ir}(z)}{\sqrt{q_{eff}}} \qquad (2)$$

The spatial phase of a Gaussian fundamental beam has a quadratic evolution with *r* with a radius of curvature, $R_{ir}(z)$. Assuming that the harmonic emission takes place near the IR beam axis, the Gaussian intensity profile can be approximated by a parabolic profile (see methods) leading to anatomic phase that has also a quadratic evolution with *r* with radius of curvature:

$$R_{atom} = \frac{k_q W_{ir}^2(z)}{4\alpha_q I_{ir}} \qquad (3)$$

with $k_q$ the $q^{th}$ harmonic wave vector. The spatial phase of the XUV beam in the emission plane is then also quadratic with *r*:

$$\varphi_q(r,z) = \frac{k_q r^2}{2}\left(\frac{1}{R_{ir}(z)} + \frac{1}{R_{atom}(z)}\right) = \frac{k_q r^2}{2R(z)} \qquad (4)$$

Under these assumptions, the XUV beam has all the characteristics (intensity profile and wavefront) of a Gaussian beam. The radii of curvature $R_{ir}$ and $R_{atom}$ can have comparable values when $\alpha_q$ is small (17). With $\alpha_q$ positive, $R_{atom}$ is positive while $R_{ir}$ can either be negative (for z<0) or positive (z>0). The curvature of the XUV wavefront in the generating medium can therefore be controlled with the jet position, $z_{jet}$. For the long quantum path or the cutoff emission where $\alpha_q$ is large (17), the impact of the atomic phase is often predominant and leads to XUV beams diverging in and after the generating medium. For the plateau harmonics emitted via the short quantum path that are considered in this paper, $\alpha_q$ is smaller and the XUV wavefront changes strongly with $z_{jet}$. As the spatial profile and wavefront are known in the generating plane, the spatial characteristics of these XUV Gaussian beams are known (24) for every longitudinal positions, *z* (see methods).

**Simulated harmonic beam during propagation.**

To simulate our experiment, the IR fundamental beam centered at λ=800 nm has a minimum beam waist $W_{0\_ir}$ = 83 μm located at *z* = 0. Harmonics are generated in neon with a peak intensity of 5.6x10$^{14}$ W/cm² at focus. We take into account that $\alpha_q$ increases with the harmonic order (21, 25-28) and consider a linear evolution from 5x10$^{-14}$ cm²/W for harmonic 29 to a constant value of 13x10$^{-14}$ cm²/W for harmonic 39 and above. The effective non linearity coefficient is $q_{eff}$ = 4.7 (see methods).

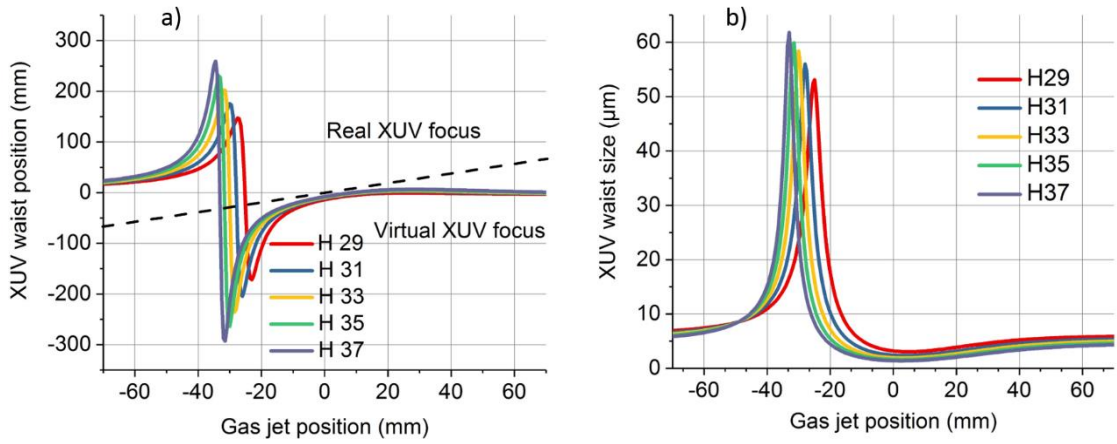

Figure 2: (a) Calculated positions of the XUV beam waists, $z_{0\_XUV}$, relative to the IR focus and (b) waists size of the harmonic beams, $W_{0\_XUV}$, as a function of the gas jet position. The IR focus is at z = 0 and negative positions imply a jet located before the IR focus.

Figure 2 shows the XUV waist positions relative to the IR focus, $z_{0\_XUV}$ (Fig 2. a) and the waist size $W_{0\_XUV}$ (Fig. 2. b) for several harmonics as a function of the gas jet position. It shows that the XUV beam waists change strongly with the gas jet position and harmonic order. The harmonic foci are seldom located at the position of the IR focus and can be shifted by almost 30 cm from it. All harmonics exhibit a similar qualitative evolution but each curve is shifted vs jet position. The XUV waists positions differ significantly from one harmonic to the next as studied in (7) and we observe that this shift is strong for all jet positions. The shift between the harmonic waists positions is large as compared to the IR Rayleigh length ($Z_{r\_ir}$ = 2.8 cm) but also as compared to the XUV beams Rayleigh lengths. Figure 2 (b) shows that the XUV beam waist, $W_{0\_xuv}$, can change with $z_{jet}$ between 3 μm and 60 μm leading to Rayleigh lengths between 1 mm ($W_{0\_xuv}$ = 3 μm for H29) and 55 cm ($W_{0\_xuv}$ = 62 μm for H37). This implies that when these harmonics are refocused on a target, their foci positions can differ significantly. Therefore, the intensity and spectral width of the XUV irradiation changes locally near each harmonic focus (7, 9, 19). The harmonic dephasing is also modified by the Gouy phase shift occurring near focus and the foci shift can modify strongly the XUV temporal profile.

In figure 2. a, the dashed line indicates the gas jet position. Below this line the XUV beams waists, located before the generating medium, are virtual. Above this line, the XUV waists located after the medium are real. Therefore, some harmonics can be emitted as diverging beams, as commonly assumed, but also as converging beams leading to an XUV beam focused after the generating medium. To the best of our knowledge, this mirrorless focusing was never observed before.

Now, we calculate the beam sizes in the far field and compare them with experimental results. When the XUV beam waist is large, the beam divergence is small leading to a small XUV beam size in the far field and vice-versa. The XUV beam size calculated 2.9 m after the IR focus is plotted in figure 3 and its evolution mimics qualitatively our experimental observations. It shows a clear minimum for $z_{jet}$ = - 25 to -35 mm followed by a maximum for $z_{jet}$ = 5 to 10 mm with a maximum beam size that increases with the harmonic order. The minimum beam size is smaller than in our measurement (0.5 mm vs 1.5 mm). It has a constant value for all harmonics and is shifted with the harmonic order as observed experimentally. If we consider that $\alpha_q$ changes with the IR peak intensity and therefore with $z$, the maxima and minima positions are shifted but the global trend is preserved (see supplementary). These results show that the XUV beam characteristics are strongly affected by the generating conditions because they define the XUV wavefront in the generating medium and the XUV wavefront evolves dramatically with $z_{jet}$.

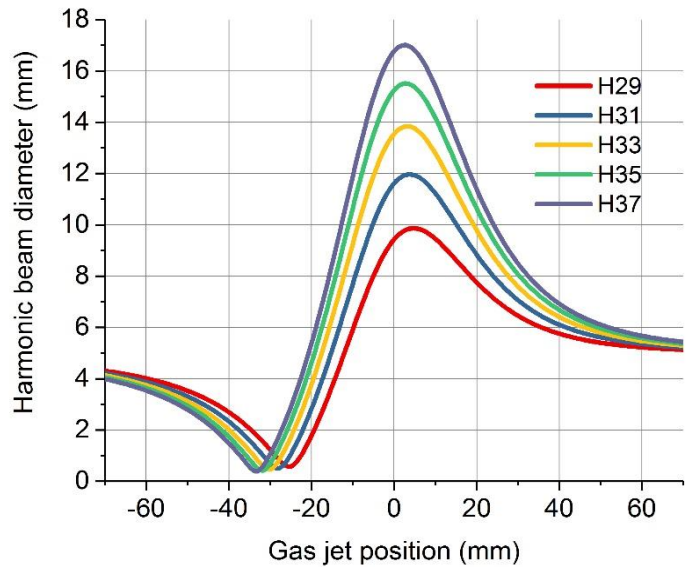

*Figure 3: Calculated diameters (FWHM) of the Gaussian XUV beams at a distance of 2.9 m after the IR focus position.*

Overall, our measurements are in good qualitative agreement with the simulations and show how the spatial properties of the XUV beams can be finely controlled. These results demonstrate the intriguing possibility of focusing the harmonics after the generating medium without any optics.

**XUV mirrorless focusing.**

To further study the generation of harmonic beams focused via XUV wavefront control, we performed HHG with a spatially chirped fundamental beam after transmitting the fundamental beam through a glass wedge (29, 30). This reduces the intensity at focus and creates a spatial asymmetry in the spectral distribution of the fundamental beam (31, 32). This asymmetry, acting as a probe inside the generation medium, changes spatially the central wavelength of the fundamental beam and of the emitted harmonics. The orientation of this spatial chirp was parallel to the entrance slit of the XUV spectrometer. The upper part of the IR beam was blue shifted and the lower part was red shifted for all longitudinal $z$ positions between $z$ = -40 mm and $z$ = +40 mm (see supplementary) and the chirp was hardly measurable afterward. In the generating medium, harmonics emitted from the top of the IR beam are therefore blue shifted and harmonics emitted from the bottom of the IR beam are red-shifted. In the far field, the spatio-spectral distribution for each harmonic is recorded for several positions of the generating medium. The spatio-spectral profiles for harmonic 29 are displayed in figure 4 for three jet positions. This beam exhibits a spatial chirp that changes with $z_{jet}$.

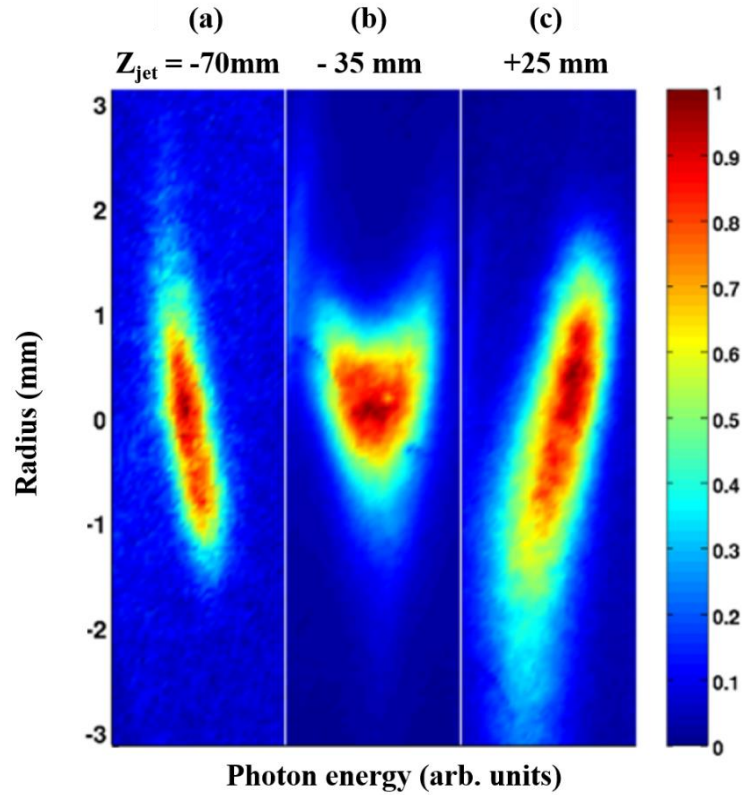

Figure 4: Far field spatially resolved normalized spectra of the harmonic 29 generated in neon for three positions of the generating medium (a) $z_{jet}$ = -70 mm, (b) $z_{jet}$ = -35 mm and (c) $z_{jet}$ = +25 mm. HHG is performed with a fundamental beam that is spatially chirped in the generating medium. The tilt of the spatially resolved harmonic spectra is the signature of a spatial chirp in the XUV beam in the far field.

For a gas jet located 25 mm after the IR focus, the XUV beam has a spatial chirp with the same orientation as the IR spatial chirp in the generating medium (upper part blue shifted). This is expected if the XUV beam is emitted as a diverging beam as the upper part of the XUV beam in the emission plane is then also the upper part in the detection plane. On the contrary, we observe a chirp orientation inversion when the gas jet is located 70 mm before the IR focus. This implies that the upper part of the beam in the generating medium is the lower part of the beam in the far field. Therefore, the upper part and lower part of the beam overlap somewhere in between these two planes confirming the mirrorless focusing of the XUV beam.

The orientation of the spatial chirp in the far field is not defined for $z_{jet}$ = -35 mm. This occurs when the XUV focus is located at the position of the gas jet (within less than the XUV Rayleigh length) since a perturbation spatially localized at the XUV focus is spatially delocalized in the far field. The XUV spatial chirps have the same orientation at $z_{jet}$ = 0 and $z_{jet}$ = +25 mm implying that the XUV beam is already diverging in the jet at $z_{jet}$ = 0. This shows directly that the XUV focus is not overlapped with the IR focus at $z_{jet}$ = 0.

Our experiment shows also that the positions of the XUV foci depend on harmonic order. The measured spatial chirps of harmonics 29 to 45 are shown in figure 5 for three jet positions.

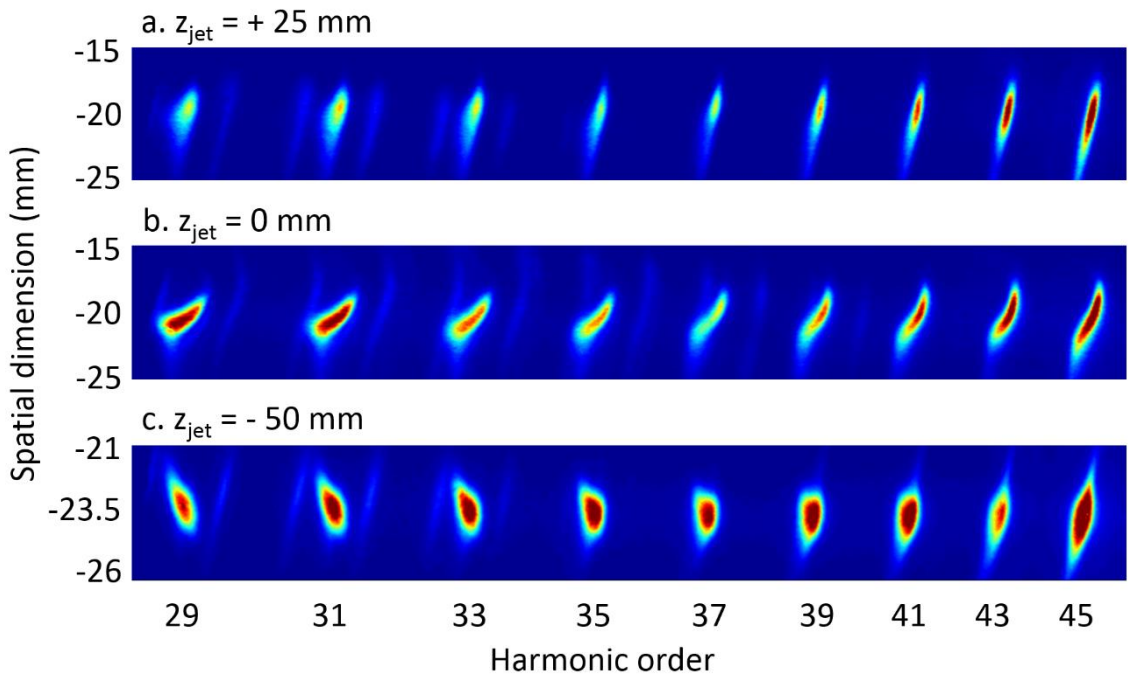

Figure 5: Far field spatially resolved high-order harmonic spectra obtained with a spatially chirped IR beam. The neon jet is located at various longitudinal positions (a) 25 mm after IR focus, (b) at IR focus and (c) 50 mm before IR focus. The harmonic orders vary between 29 and 45 for the first order of diffraction of the XUV grating and we observe the second orders of diffraction for harmonics 59 and higher. The tilt of the harmonic beams is the signature of a spatial chirp in the XUV beams in the far field. For the position $z_{jet}$ = -50mm where the XUV beam is small, the spatial scale has been expanded.

The figure 5 shows that the orientation and magnitude of the far-field XUV spatial chirp change continuously with the harmonic order. For harmonic orders larger than 37, the spatial chirp has the same orientation as in the generating medium for all gas jet positions. These harmonics are always diverging in and after the generation medium and the XUV beam waists are virtual. For $z_{jet}$ = -50 mm the spatial chirp is inverted for low order harmonics (q < 35) meaning that these XUV beams are focused after the generating medium. The far field spatial chirp of the XUV beam disappears when the XUV focus is located onto the jet (within the XUV Rayleigh range) and its orientation shows if the waists are real or virtual. Observing an inversion of the spatial chirp with harmonic order therefore confirms that for $z_{jet}$ = -50 mm, the foci of harmonics 29 to 35 and 41 to 65 are located on opposite sides of the generating medium and are separated by several XUV Rayleigh lengths. If these harmonic beams were refocused on a target, the XUV beams would not overlap significantly near their foci.

## DISCUSSION

These results, obtained under standard HHG conditions, question how broadband XUV radiation should be used in attoscience experiments since the harmonic foci relative shift are large for all jet positions. This shift reduces locally the XUV bandwidth near each focus where a specific harmonic is enhanced. A first solution to maintain the XUV bandwidth is to image the HHG medium onto the target.

In this image plane, each harmonic intensity is lower than at the harmonic focus but all the harmonics have similar spatial profiles ensuring broadband irradiation. This limits the thickness of the target as the spatial overlap is reduced before/after the image plane. Another solution is to reduce strongly the XUV beam size in the far field with an iris before refocusing the beam on an experiment (5, 33). The beam clipping enlarges the XUV Rayleigh lengths and for strong clipping the Rayleigh lengths can get larger than the relative foci shift. Both options reduce the useful XUV intensity and it would be interesting to control the XUV beam shape to obtain simultaneously broadband irradiation and high XUV intensities. This can be done with controlled mirrorless focusing as shown in the next paragraph. At the opposite, the XUV foci shift could be used to perform spectral filtering of harmonic sources by locating a pinhole near the harmonic foci (34). The pinhole can transmit a specific harmonic group while reducing the transmission of other harmonics and of the fundamental beam. This provides an efficient spectral shaping with limited temporal stretching or even temporal compression when the attosecond pulses are strongly chirped before spectral selection (35).

Mirrorless focusing has never been reported in standard HHG setup but this effect is robust and likely universal when HHG occurs in a gas medium located before the laser focus. The control of the harmonic spatial characteristics via spatial coherence is demonstrated here with a beam having a Gaussian spatial profile. It can be extended by controlling spatial profile and wavefront of the fundamental beam. Using a fundamental beam with a radially flat-top intensity profile (36) and a spherical wavefront would for instance make the mirrorless focusing mostly independent on harmonic orders and laser intensity. Several harmonics could then be tightly refocused without any focusing optics over a large bandwidth compatible with broadband and intense attosecond irradiation. By considering actual record values of XUV energy of 10 µJ per isolated attosecond pulse (37) and assuming a focusing on a 20-micron waist with duration of 200 attosecond, we estimate that XUV intensities larger than $10^{15}$ W/cm$^2$ can be achieved with mirrorless focusing. Higher XUV intensities should even be accessible with higher energy fundamental lasers (38). With precise wavefront control of an IR fundamental beam, this mirrorless focusing can even be applied to focus harmonics generated in the soft X-ray domain (P3) that are harder to control spatially.

This study shows therefore that the spatial characteristics of XUV high-order harmonic beams generated in a thin medium can be controlled directly via a shaping of the XUV source wavefront and spatial distribution in the emission medium. We present a Gaussian model to infer the XUV beams spatial properties and show how the position of the XUV beam waists and the propagation characteristics of the XUV beams depend on both the harmonic order and the conditions of generation. By performing HHG in neon gas with a spatially filtered, wavefront corrected, high energy femtosecond laser, we demonstrate that the longitudinal gas medium displacement allows controlling the XUV beams properties and that plateau harmonic beams can be focused directly after the generating medium without any focusing optics. The XUV beam waists can be virtual for some harmonics and real for some others under the same generating conditions. The observed order-dependent foci shift implies that the harmonic foci are, in general, separated and we show experimentally that their separation can be as large as several Rayleigh lengths of the XUV beams. We discuss how this relative foci shift is compatible with attosecond science and propose ways to use mirrorless focusing to perform XUV spectral filtering or achieve high XUV intensities.

## MATERIAL AND METHODS
**Experimental setup.**

We used a high energy femtosecond fundamental IR (centered at $\lambda_0$ = 800 nm) pulse and the laser beam was spatially filtered before compression. Spatial filtering was performed under vacuum with a 250 µm pinhole after clipping the incoming beam with an iris and focusing it with a f = 1.3 m lens. The beam was afterward collimated with a spherical mirror located 1.5 m after the pinhole. The overall transmission of the spatial filtering was 50% and output pulses with maximum energies of 50 mJ after compression could be used. When the spatial chirp was introduced the pulse energy was limited to 5 mJ to limit non-linear effects in the wedge.

The IR wavefront is corrected after compression under vacuum by a deformable mirror (High Power Active Mirror, Hipao, ISP Systems) that was specifically designed for high energy ultrashort lasers. This active mirror is equipped with a thick glass membrane with multilayer dielectric coating that allows smooth correction and stable operation. The corrected IR wavefront, measured with an HASO (Imagine Optics), showed a residual error smaller than 8 nm rms ($\lambda$/100) and the Strehl ratio was larger than 0.95. Controlling the wavefront with such a high precision is crucial since it ensures good XUV spatial coherence even when the harmonics are generated far from the IR focus position. Wavefront control was also used here to pre-compensate for the astigmatism induced by the spherical mirror used to focus the IR beam in the gas jet.

After corrections, we observed that the intensity profile of the fundamental beam was close to a Gaussian beam in the far field and near focus with a small asymmetry. The measured laser beam factor ($M^2$) was slightly different in the horizontal and vertical direction (see supplementary) and we used a value of $M^2$=1.04 in the simulations. We measured a beam profile and size evolution with *z* that was close to a perfect Gaussian beam (see supplementary) and we assumed in simulations that the fundamental beam propagates (waist and wavefront dependence with *z*) like a Gaussian beam.

We use a 40 fs fundamental pulse with energy of ~ 5 mJ to generate high-order harmonics by focusing the input beam ($W_x$ = 6.7 mm and $W_y$ = 7.6 mm) truncated by a 20 mm diameter iris with a f = 2m spherical mirror to a focal spot size of $W_0 \approx$ 83 µm. The iris transmits most of the beam power and improves the beam symmetry near focus. HHG occurs at 10 Hz in a pulsed neon jet with 200 µm nozzle diameter that could be moved by 7.5 cm before and after the IR focus. We analyze the spatial profile of each harmonic in the far field with a flat field spectrometer. It is equipped with a 500 µm wide entrance slit, a grazing incidence grating with 1200 grooves / mm (Hitachi aberration corrected concave grating) and a 40 mm diameter Micro channel plate (MCP) detector that was located 2.9 m after the IR focus position.

**Analytical method.**

In this paper, we neglect propagation in the medium by considering an infinitely thin medium. Experimentally, this implies a loose focusing configuration and the use of conditions where the reshaping of the fundamental beam does not occur (6, 39, 40). For a medium that is only smaller than the IR beam confocal parameter, longitudinal phase matching can induce an additional amplitude pre-factor (22) that can slightly modify the XUV beam spatial profile in the generating medium. Longitudinal phase matching/mismatch in thick media can also lead to the selection of a subset of XUV wavectors that are already existing with the thin medium considered here. It can however not lead to the creation of new XUV wavevectors and thereby the observations described here with a thin medium should mostly hold with extended media. In general, our approach can be extended to more complex propagation situation but this is beyond the scope of this paper that focuses on the physics that allows controlling the spatial characteristics of coherent XUV beams obtained via HHG in gases.

In our model, we consider a single peak intensity even if the peak intensity changes with time in a real pulse. This approximation was done in order to mimic conditions under which a single attosecond pulse is emitted. Indeed, isolated attosecond pulse emission occurs at a specific time where the intensity is fixed. Averaging over several intensities could lead to a better agreement with experiments and the value of the minimum would likely be bigger than in this simulations after averaging over several intensities. It could also lead to more limited apparent beam evolution and can reduce the apparent foci shift. However, we stress that when a single attosecond pulse is generated, the emission time correspond to a single peak intensity and therefore our simulations, providing the beam shape at a specific time, are relevant to highlight the underlying physics.

The atomic phase of the model dipole is :

$$\varphi_{atom}(r) = -\alpha_q\, I(z_{jet})\, \exp(-2\, r^2/W_{ir}(z_{jet})^2) \qquad (5)$$

It is simplified by performing a Taylor expansion of the Gaussian IR intensity profile near the laser axis that leads to a quadratic phase:

$$\varphi_{atom}(r) = k_q\, r^2/2R_{atom} \qquad (6)$$

With:

$$R_{atom} = \pi W_{ir}^4 / 2\alpha_q I_0 W_{0\_ir}^2 \lambda_q \qquad (7)$$

where $\lambda_q$ is the harmonic wavelength, $W_{0\_ir}$ the IR beam waist size at focus, $W_{ir}$ the IR beam size at the jet position and $I_0$ the peak intensity at focus.

Since the IR wavefront is also quadratic with radius of curvature $R_{ir}$, the emitted XUV wavefront is spherical with a radius of curvature $R$ given by $1/R = 1/R_{ir} + 1/R_{atom}$

The XUV intensity profile is Gaussian in the generating medium and its size is linked to the IR beam size via the effective order of non-linearity $q_{eff}$ (eq. 2). With a Gaussian spatial profile and a spherical wavefront, the XUV beams have all the characteristics of regular Gaussian beams and from standard Gaussian propagation law (24), we obtain analytical formulas for both the XUV focus position, $z_{0\_XUV}$ as compared to the IR focus and the harmonic waist size, $W_{0\_XUV}$:

$$z_{0\_xuv} = z_{jet} - R / (1 + R^2\, \lambda_q^2\, q_{eff}^2 / (\pi^2 W_{ir}^4)) \qquad (8)$$

$$W_{0\_xuv} = [(R^2\, \lambda_q^2\, q_{eff}^2 / (\pi^2 W_{ir}^2))/(1 + R^2\, \lambda_q^2\, q_{eff}^2 / (\pi^2 W_{ir}^4))]^{1/2} \qquad (9)$$

The displacement of the XUV waist position, $z_{0\_xuv}$, depends on the XUV wavefront radius of curvature, $R$, in the generating medium. When $z_{jet}$ is neglected, this displacement is maximum for

$$R^2\, \lambda^2\, q_{eff}^2 / (\pi^2 W_{ir}^4) = 1 \qquad (10)$$

or equivalently for

$$R = -\pi W_{ir}^2 / (\lambda_q q_{eff}) \qquad (11)$$

The maximum shift of the XUV waist position is therefore

$$z_{0\_xuv} = -R/2 = (\pi W_{ir}^2) / (2 \lambda_q q_{eff}) \qquad (12)$$

that can be expressed as:

$$z_{0\_xuv} = 1/2\, z_{r\_ir} (W_{ir}/W_{0\_ir})^2\, q/q_{eff} \qquad (13)$$

This analytical expression (eq. 13) shows that the XUV beam can be focused far from the IR focus provided that $q_{eff}$ is smaller than the harmonic order (always correct in the tunneling regime) and provided that harmonic generation takes place off focus where $W_{ir}$ is significantly larger than $W_{0\_ir}$.

The size of the XUV waist, $W_{0\_xuv}$ can also be estimated at this maximum shift by combining eq. (9) and eq. (11) that leads to:

$$W_{0\_xuv} = \frac{1}{\sqrt{2}} \frac{W_{ir}}{\sqrt{q_{eff}}} \qquad (14)$$

This last formula shows that when the focus is at the maximum shift, the XUV beam waist is independent on the XUV wavelength and it is equal to the size of the XUV beam in the gas jet divided by $\sqrt{2}$. It also implies that the maximum foci shift is the XUV Rayleigh length that can be arbitrarily large for extended XUV sources. This estimate gives the waist located at the maximum distance of the generating medium and these waists are large at that position. Smaller XUV waist can be obtained closer to the generating medium (Fig1. b) and controlling the generating medium position is a way to control the XUV waist size and location between this maximum shift and a minimum waist size obtained near the IR focus.

The formula (8) and (9) also show that the largest XUV waists are obtained when the XUV foci overlap with the generating medium (obtained around z=-30 mm in figure 3). In return, this shows that the maximum waist size is similar for all harmonics provided that $q_{eff}$ is approximately constant. This implies that the minimum beam size in the far field must be nearly identical for all harmonics as observed experimentally.

We note that the harmonic foci shift depend on the fundamental wavelength that impacts strongly $q/q_{eff}$ but also the value of $\alpha_q$ since its asymptotic value ($\alpha$ value in the cutoff that represents a scaling factor in the mirrorless focusing) is proportional to the third power of the laser frequency (22, 28). Mid-IR and far-IR fundamental lasers are now used to obtain broadband harmonic plateaus and we anticipate that the cutoff harmonic will be focused very far from the lowest plateau harmonics and the harmonic foci shift should be very pronounced with these lasers. For a specific fundamental wavelength, the relative foci shift between consecutive harmonics depends on the evolution of the $\alpha_q$ parameter between each harmonic. This parameter is close to zero near $I_p$ for short trajectories harmonic and close an asymptotic value mostly defined by the fundamental wavelength near the cutoff. Therefore, for a given fundamental wavelength the $\alpha_q$ change between each harmonic is more pronounced for a narrower plateau. Thus the foci shift, shown here when neon atoms are used for

HHG, should also be observable with other generating gases like Xe, Kr or Ar (7) that leads to narrower plateaus but faster evolution of the $\alpha_q$ term between adjacent harmonics.

**SUPPLEMENTARY MATERIALS**

Supplementary material is available for this article.
section S1. Experimental setup.
section S2. Characterization of the IR spatial chirp near focus.
section S3. Evolution of the HHG efficiency with the jet longitudinal position.
section S4. Evolution of the beam size for higher order harmonics.
section S5. Impact of $q_{eff}$ and $\alpha$ on the numerical results.
section S6. Impact of the longitudinal evolution of $\alpha$.
section S7. Normalized foci shift.
section S8. Simulations with non-Gaussian XUV beams.

fig. S1. Spatial profile of the fundamental beam.
fig. S2. Experimental setup.
fig. S3. Characterization of the spatial chirp near the focus of the IR beam.
fig. S4. Integrated XUV emission.
fig. S5. Diameter of the XUV beam in the far field.
fig. S6. Parametric evolution of the XUV beam size.
fig. S7. Influence of the $\alpha$ parameter.
fig. S8. Beam size evolution for several peak intensities.
fig. S9. Normalized foci shift.
fig. S10. Non Gaussian XUV beam size evolution.

References:

1. G. Sansone, L. Poletto, M. Nisoli, "High-energy attosecond light sources", Nat. Photonics **5**, 655 - 664 (2011)

2. C. M. Heyl, H. Coudert-Alteirac, M. Miranda, M. Louisy, K. Kovacs, V. Tosa, E. Balogh, K. Varjú, A. L'Huillier, A. Couairon, and C. l. Arnold "Scale-invariant nonlinear optics in gases ", Optica **3**, 75 - 81 (2016)

3. T. Popmintchev, M. Chen, P. Arpin, M. M. Murnane and H. C. Kapteyn, "The attosecond nonlinear optics of bright coherent X-ray generation", Nat. Photon. **4**, 822 - 832 (2010)

4. M. Nisoli, E. Priori, G. Sansone, S. Stagira, G. Cerullo, and S. De Silvestri, C. Altucci, R. Bruzzese and C. de Lisio, P. Villoresi, L. Poletto, M. Pascolini, and G. Tondello, "High-Brightness High-Order Harmonic Generation", Phys. Rev. Lett. **88**, 033902 - 1 - 4 (2002)

5. M. B. Gaarde and K. J. Schafer, "Generating single attosecond pulses via spatial Filtering", Opt. Lett. **31**, 3188 - 3190 (2006)

6. M. B. Gaarde, Tate, J. L. & Schafer, K. J. "Macroscopic aspects of attosecond pulse Generation", J. Phys. B **41**, 132001 1 - 26 (2008)

7. E. Frumker, G. G. Paulus, H. Niikura, A. Naumov, D. M. Villeneuve, and P. B. Corkum, "Order-dependent structure of high harmonic wavefronts", Opt. Express. **20**, 13870 - 13877 (2012)

8. A. Dubrouil, O. Hort, F. Catoire, D. Descamps, S. Petit, E. Mével, V. Strelkov and E. Constant, "Spatio–spectral structures in high-order harmonic beams generated with Terawatt 10-fs pulses", Nat. Comm. **5**, 4637 1 – 8 (2014)

9. D. T. Lloyd, K. O'Keefe, P. N. Anderson, & S. Hooker, "Gaussian-Schell analysis of the transverse spatial properties of high-harmonic beams", Scientific report **6**, 30504 1 – 9 (2016)

10. Z. Li, G. Brown, D. H. Ko, F. Kong, L. Arissian, and P. B. Corkum, "Perturbative High Harmonic Wavefront Control", Phys. Rev. Lett. **118**, 033905 1- 5 (2017)

11. M. Sivis, M. Taucer, G. Vampa, K. Johnston, A. Staudte, A. Y. Naumov, D. M. Villeneuve, C. Ropers and P. B. Corkum, "Tailored semiconductors for high-harmonic optoelectronics", Science **357**, 303 – 306 (2017)

12. N. Dudovich, J. L. Tate, Y. Mairesse, D. M. Villeneuve, P. B. Corkum, and M. B. Gaarde, "Subcycle spatial mapping of recollision dynamics", Phys. Rev. A **80**, 011806R 1 – 4 (2009)

13. P. Tzallas, D. Charalambidis, N. A. Papadogiannis, K. Witte & G. D. Tsakiris, "Direct observation of attosecond light bunching", Nature **426**, 267 - 271 (2003)

14. T. Sekikawa, A. Kosuge, T. Kanai & S. Watanabe, "Nonlinear optics in the extreme ultraviolet", Nature **432**, 605 - 608 (2004)

15. T. Okino, Y. Furukawa, T. Shimizu, Y. Nabekawa, K. Yamanouchi and K. Midorikawa, "Nonlinear Fourier transformation spectroscopy of small molecules with intense attosecond pulse train", J. Phys. B: At. Mol. Opt. Phys. **47,** 124007 1-23 (2014)

16. B. Manschwetus, L. Rading, F. Campi, S. Maclot, H. Coudert-Alteirac, J. Lahl, H. Wikmark, P. Rudawski, C. M. Heyl, B. Farkas, T. Mohamed, A. L'Huillier, and P. Johnsson, "Two-photon double ionization of neon using an intense attosecond pulse train", Phys. Rev. A **93**, 061402(R) 1-5 (2016)

17. M. Lewenstein, P. Salieres, and A. L'Huillier, "Phase of the atomic polarization in high-order harmonic generation", Phys. Rev. A **52**, 4747-4754 (1995)

18. M. B. Gaarde, F. Salin, E. Constant, Ph. Balcou, K. J. Schafer, K. C. Kulander, and A. L'Huillier, "Spatio-temporal separation of high harmonic radiation into two quantum path components", Phys. Rev. A **59**, 1367-1373 (1999)

19. V. T. Platonenko and V. V. Strelkov, "Single attosecond soft-x-ray pulse generated with a limited laser beam", J. Opt. Soc. Am. B **16**, 435 - 440 (1999)

20. E. Constant, D. Garzella, P. Breger, E. Mével, Ch. Dorrer, C. Le Blanc, F. Salin, and P. Agostini, "Optimizing High Harmonic Generation in Absorbing Gases: Model and Experiment", Phys. Rev. Lett. **82**, 1668 - 1671 (1999)

21. S. Carlström, J. Preclíková, E. Lorek, E. W. Larsen, C. M. Heyl, D. Paleček, D. Zigmantas, K. J Schafer, M. B Gaarde and J. Mauritsson, "Spatially and spectrally resolved quantum path interferences with chirped driving pulses", New Jour. Phys. **18**, 123032 1-17 (2016)

22. F. Catoire, A. Ferré, O. Hort, A. Dubrouil, L. Quintard, D. Descamps, S. Petit, F. Burgy, E. Mével, Y. Mairesse, and E. Constant, "Complex structure of spatially resolved high-order-harmonic spectra", Phys. Rev. A **94**, 063401 1-14 (2016)

23. L. Rego, J. S. Román, A. Picón, L. Plaja, and C. Hernández-García, "Nonperturbative Twist in the Generation of Extreme-Ultraviolet Vortex Beams", Phys. Rev. Lett. **117**, 163202 1-6 (2016)

24. H. Kogelnik and T. Li, "Laser Beams and Resonators", Appl. Opt. **5**, 1550 - 1567 (1966)

25. M. B. Gaarde and K. J. Schafer, "Quantum path distributions for high-order harmonics in rare gas atoms", Phys. Rev. A **65**, 031406R 1-4 (2002)

26. C. Corsi, A. Pirri, E. Sali, A. Tortora, and M. Bellini, "Direct Interferometric Measurement of the Atomic Dipole Phase in High-Order Harmonic Generation", Phys. Rev. Lett. **97**, 023901 1-4 (2006)

27. L. He, P. Lan, Q. Zhang, C. Zhai, F. Wang, W. Shi, and P. Lu "Spectrally resolved spatiotemporal features of quantum paths in high-order-harmonic generation", Phys. Rev. A **92**, 043403 1-9 (2015)

28. M. Khokhlova and V. Strelkov 2016, "Phase properties of the cutoff high-order harmonics", Phys. Rev. A **93**, 043416 1-7 (2016)

29. H. Vincenti and F. Quéré, "Attosecond Lighthouses: How To Use Spatiotemporally Coupled Light Fields To Generate Isolated Attosecond Pulses", Phys. Rev. Lett. **108**, 113904 1-5 (2012)

30. F. Quéré, H. Vincenti, A. Borot, S. Monchocé, T. J. Hammond, K. T. Kim, J A Wheeler, C. Zhang, T. Ruchon, T. Auguste, J. F. Hergott, D. M. Villeneuve, P. B. Corkum and R. Lopez-Martens "Applications of ultrafast wavefront rotation in highly nonlinear optics", J. Phys. B **47**, 124004 1-23 (2014)

31. Z. Guang, M. Rhodes, and R. Trebino, "Measurement of the ultrafast lighthouse effect using a complete spatiotemporal pulse-characterization technique", J. Opt. Soc. Am. B **33**, 1955 - 1962 (2016)

32. C. Hernández-García, A. Jaron-Becker, D. D. Hickstein, A. Becker and Ch. G. Durfee, "High-order-harmonic generation driven by pulses with angular spatial chirp"" Phys. Rev. A **93**, 023825 1-7 (2016)

33. M. Hentschel, R. Kienberger, Ch. Spielmann, G. A. Reider, N. Milosevic, T. Brabec, P. Corkum, U. Heinzmann, M. Drescher, & F. Krausz, "Attosecond metrology", Nature **414**, 509-513 (2001)

34. V. T. Platonenko, V. V. Strelkov, "Generation of a single attosecond x-ray pulse", Quantum Electronics **28**, 749–753 (1998)

35. Y. Mairesse, A. de Bohan, L. J. Frasinski, H. Merdji, L. C. Dinu, P. Monchicourt, P. Breger, M. Kovačev, R. Taïeb, B. Carré, H. G. Muller, P. Agostini, P. Salières, "Attosecond synchronization of high harmonic soft X-rays", Science **302**, 1540 - 1543 (2003)

36. A. Dubrouil, Y. Mairesse, B. Fabre, D. Descamps, S. Petit, E. Mevel and E. Constant, "Controlling high harmonics generation by spatial shaping of high energy femtosecond beam", Opt. lett. **36**, 2486-2488 (2011)

37. E. J. Takahashi, P. Lan, O. D. Mucke, Y. Nabekawa & K. Midorikawa, "Attosecond nonlinear optics using gigawatt-scale isolated attosecond pulses", Nat. Comm. **4,** 2691 1-9 (2013)

38. S. Kuhn et al, "The ELI-ALPS facility: the next generation of attosecond sources", J. Phys. B At. Mol. Opt. Phys. **50**, 132002 1-39 (2017)

39. W. Holgado,_ B. Alonso, J. San Romanand I. J. Sola, "Temporal and spectral structure of the

infrared pulse during the high order harmonic generation", Opt. Exp. **22**, 10191-10201 (2014)

40. V. Tosa, E. Takahashi, Y. Nabekawa, and K. Midorikawa, "Generation of high-order harmonics in a self-guided beam", Phys. Rev. A **67**, 063817 1-4 (2003)

**Acknowledgements:** We acknowledge stimulating discussions with F. Lépine, V. Loriot, K. Veyrinas and C. Valentin. **Fundings:** We acknowledge funding from the region Aquitaine (Caracatto 2013-1603008 and Hipao programs), the CNRS (Pics n°PICS06038), the ANR (ANR-09-BLAN-0031-02 Attowave and ANR-16-CE30-0012 Circé), RFBR (grant N 16-02-00858). Preliminary theoretical studies showing the existence of mirror less focusing were supported by RSF (grant N 16-12-10279). **Author contributions:** L. Q., E. M., E. C., performed the experiments. E. C. and E. M. conceived the experiment and, with O. H., A. D., D. D., F. B., C. P. developed the apparatus. F. C., J. V. and V.S. developed the simulations. L. Q. and E. C. performed the data analysis. First drafting of the manuscript was carried out by E. C. All authors contributed to the final version of the manuscript. **Competing interests:** The authors declare that they have no competing interests. **Data and materials availability:** All data needed to evaluate the conclusions in the paper are present in the paper and/or the Supplementary Materials. Additional data related to this paper may be requested from the authors.

# Supplementary information for

# "Mirrorless focusing of XUV high-order harmonics"

L. Quintard[1], V. Strelkov[2], J. Vabek[1], O. Hort[1,†], A. Dubrouil[1,‡], D. Descamps[1], F. Burgy[1], C. Péjot[1], E. Mével[1], F. Catoire[1], and E. Constant[1,3].

[1]*Université de Bordeaux, CNRS, CEA, Centre Laser Intenses et Applications (CELIA), 43 rue P. Noailles, 33400 Talence, France*

[2]*A M Prokhorov General Physics Institute of Russian Academy of Sciences, 38, Vavilova Street, Moscow 119991, Russia.*
*Moscow Institute of Physics and Technology (State university), 141700 Dolgoprudny, Moscow Region, Russia.*

[3]*Univ Lyon, Université Claude Bernard Lyon 1, CNRS, Institut Lumière Matière (ILM), F-69622 Villeurbanne, France.*

† *present address: ELI Beamlines Project, Institute of physics, Czech Academy of Sciences, Na Slovance 1999/2, 182 21 Praha 8, Czech Republic*

‡ *present address : Femtoeasy, Femto Easy SAS, Batiment Sonora, parc scientifique et technologique Laseris 1 , 33114 Le Barp, France*

**S1. Experimental setup.**

The laser used in this experiment is a terawatt Ti:Sapphire laser delivering up to 100 mJ per pulse at a repetition rate of 10 Hz. The input spatial profile is initially super Gaussian and it is spatially filtered (with a transmission of 50%) to obtain a regular spatial profile with a near Gaussian intensity profile (Fig. S1).

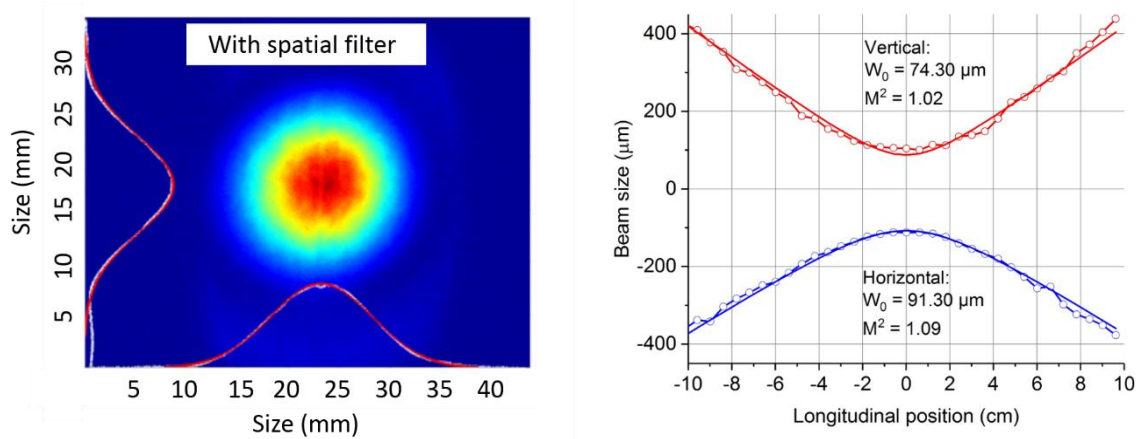

**Fig. S1: Spatial profile of the fundamental beam.** Spatial profile of the collimated IR beam after spatial filtering and measurements of the size of the focused beam near the place where harmonics are generated.

The beam propagates for almost 10 m after the spatial filtering but the spatial profile remained correct despite a small asymmetry. The beam profile is close to Gaussian with a beam size $W_x$ = 6.7 mm and $W_y$ = 7.6 mm. Before HHG the beam was clipped by a 20 mm diameter iris to remove external rings and to improve the beam symmetry.

After spatial filtering, the pulses are recompressed under vacuum down to 40 fs with a grating compressor. Afterward the short pulses propagate under vacuum for HHG and XUV beam characterization (Fig. S2).

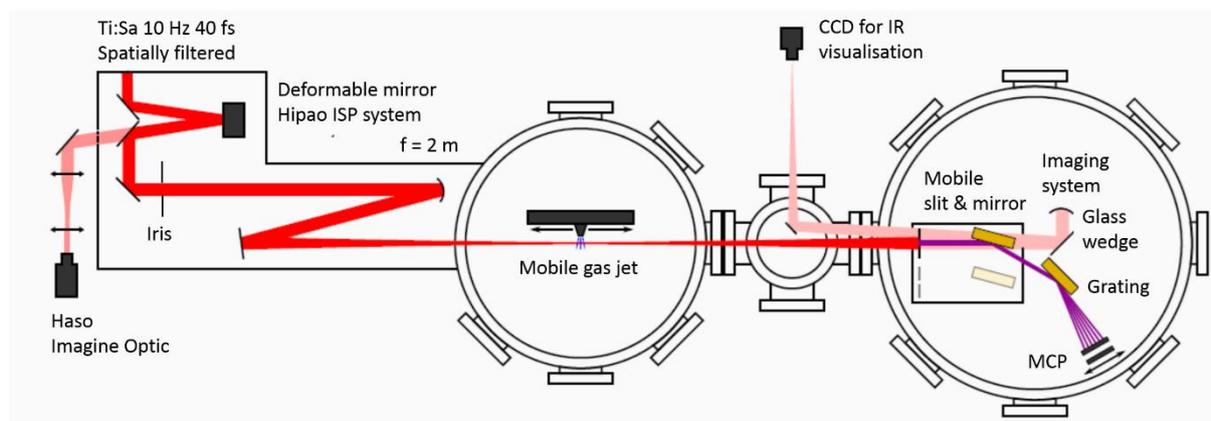

**Fig. S2: Experimental setup.** Sketch of the experimental setup that shows the deformable mirror and HASO (left), the harmonic generation chamber (center) and a part devoted to IR and XUV beam characterization (achromatic imaging system and flat field XUV spectrometer).

In order to improve the IR wavefront, the beam is reflected by a deformable mirror (Hipao, ISP System) equipped with a thick multilayer dielectric coated membrane specifically designed for high energy femtosecond lasers. The corrected wavefront was measured with a Haso wavefront sensor (Imagine Optic) and showed a residual error of less than 8 nm rms. We measured Strehl ratio was higher than

0.95 in few iterations and the correction was stable for a full day of experiment. The active mirror is also used to pre-compensate for the astigmatism that is induced by the 2 m spherical mirror focusing the IR beam in the gas jet. This improves the beam spatial quality near focus.

After wavefront correction, the beam is focused with a f = 2m focal length spherical mirror down to a spot size of ~83 µm and the corresponding Rayleigh length is $Z_r$ = 2.7 cm. A pulsed mobile gas jet was used for HHG. It was located near focus and could be moved by 75 mm before and after the focus. The beam profile evolution near focus was measured with a reflective achromatic imaging system that attenuated the beam with three reflections on 45° incidence $SiO_2$ wedges before exiting the vacuum chamber where the spatial profile was recorded on a CCD camera. The imaging system allowed us to observe the IR beam profile evolution near focus at full power. It was also used to measure the exact position of the gas jet as compared to the beam focus.

**S2 Characterization of the IR spatial chirp near focus.**

For some experiments, we inserted a small angle $SiO_2$ prism after the deformable mirror and in front of the motorized iris to induce spatial chirp on the fundamental beam. This spatial chirp was characterized near focus with the above mentioned imaging system. For this measurement, we measured the IR spectrum at several heights inside the beam and for several longitudinal positions (Fig. S3). We observed that near focus, the spatial chirp has always the same orientation (31): the top part of the beam is blue shifted as compared to the bottom part of the beam. We observed that the sign of the spatial chirp of the IR beam was identical for z positions evolving from -40 mm to +40 mm and was hardly measurable afterward.

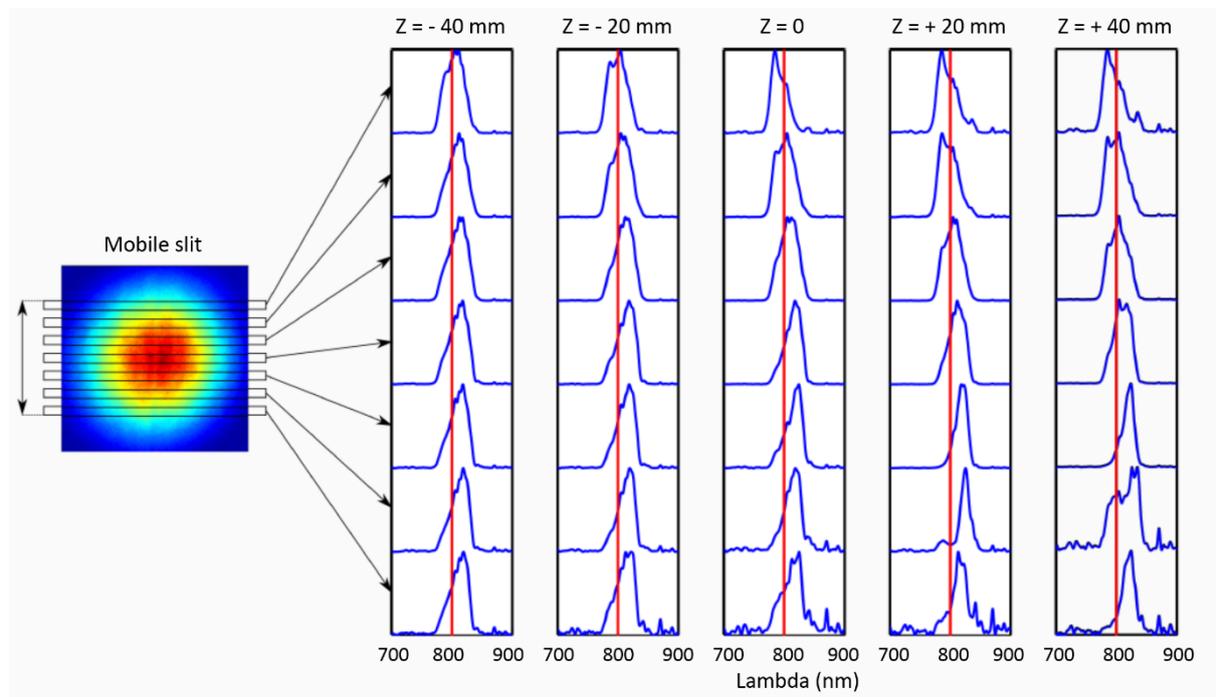

**Fig. S3: Characterization of the spatial chirp near the focus of the IR beam.** The spectrum is measured inside the image plane at several heights of a mobile slit located before a spectrometer. These measurements show that the upper part of the beam is blue shifted as compared to the lower part of the beam.

**S3 Evolution of the HHG efficiency with the jet longitudinal position.**

The fundamental beam was used to perform HHG in neon gas and the XUV beams were characterized spatially and spectrally with a flat field XUV spectrometer equipped with a 500 µm input slit. We measured the spatial characteristics of the XUV beam as a function of the gas jet position but also the relative efficiency of the HHG process (Fig.S4)

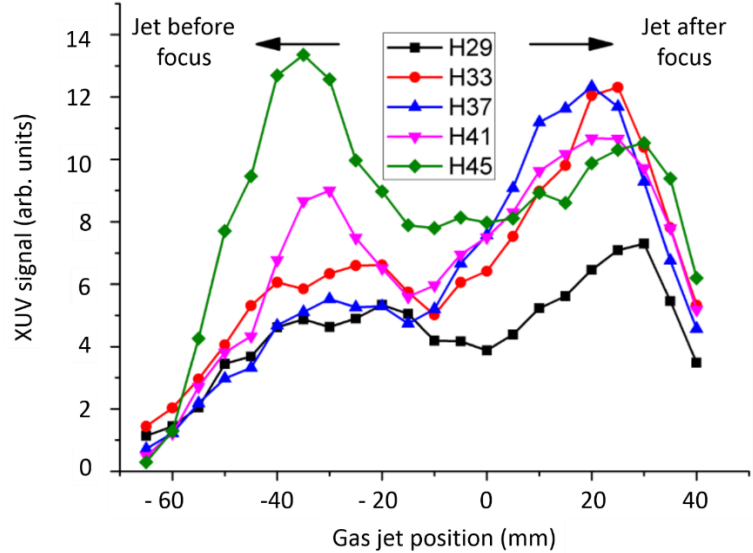

**Fig. S4: Integrated XUV emission.** Evolution of the signal emitted for each harmonic as a function of the longitudinal medium position.

The signal emitted for each harmonic was estimated from the recorded signal after correction of the slit transmission. This transmission depends on the beam size on the input slit and evolves as:

$$T = \mathrm{erf}(\frac{L}{\sqrt{2}W})$$

where $L$ is the slit width (500 µm) and $W$ the beam size on the slit (approximated here by scaling the size measured on the detector and by considering that all XUV foci were located at the position of the IR focus). We observe that the signal remains at a high level for most of the longitudinal positions that we used in this work. This implies that the XUV beam spatial coherent control demonstrated here is compatible with efficient HHG.

**S4 Evolution of the beam size for higher order harmonics.**

The XUV beam divergence showed a strong evolution with the longitudinal position of the jet as described for the harmonics 29 to 37 in the main text (Fig. S5.a). We also recorded the evolution of the beam size for higher order harmonics and observed also a regular evolution (Fig. S5b). In that case, the wavelength dependence on the beam size was less pronounced than for lower harmonics. The measured maximum beam size evolution was even opposite to the dependence observed with low order harmonics. Above order 39, we observed that the higher the harmonic order, the smaller the maximum beam size. We could not observe any minimum of the beam size for those high harmonics. Based on our studies on lower order harmonics where a minimum in the beam size was obtained only when the harmonics were converging in and after the gas jet, we conclude that these harmonics are always diverging after the gas jet. These two observations point toward the fact that the harmonics with orders higher than 39 behave like cutoff harmonics where the alpha term does not significantly change with the harmonic order and where it is large enough to ensure that harmonics are always diverging after the generating medium.

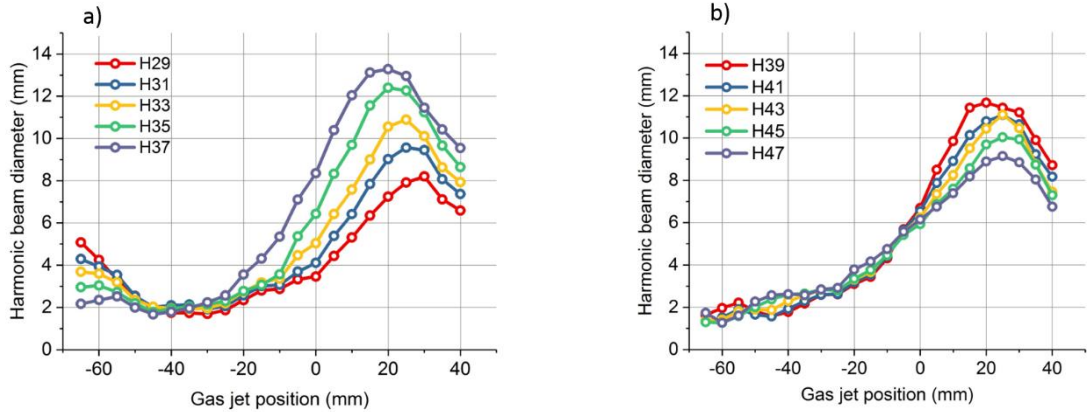

**Fig.S5: Diameter of the XUV beam in the far field.** (a) Evolution of the beam size on the detector for the harmonics presented in the main text (order 29 to 37) and (b) for higher order harmonics (orders 39 to 47).

**S5 Impact of $q_{eff}$ and $\alpha$ on the numerical results.**

The theoretical model developed in this paper allows easy manipulation of the important parameters used in this paper such as $\alpha$ and $q_{eff}$. When alpha was taken as constant for all harmonics (Fig. S6 with $\alpha$ = 5 and $q_{eff}$ = 4.7 for all harmonics), we observe that the lower harmonics have a larger maximum beam size in contrast with our experimental observations on harmonics with orders in the range 29 to 37. This evolution corresponds to the evolution observed for the highest harmonics (Fig. S5 b). For these harmonics the $\alpha$ parameter can effectively be constant if they behave like cutoff harmonics. This is consistent with a recent publication showing that large group of harmonics can behave like cutoff harmonics (28). We also checked the influence of $q_{eff}$ that can impact the beam size in the far field. This parameter $q_{eff}$ defines the harmonic source size and it is expected that $q_{eff}$ increases with the harmonic order since the higher order harmonic are generated in a volume that is more confined near the laser axis than the lower order harmonics. We performed simulations with $q_{eff}$ increasing linearly by 1 with the harmonic order ($q_{eff}$ = 4.7 for h29 and $\alpha$ = 5 for all harmonics) as shown in Fig. S6.b. We observe that increasing $q_{eff}$ tends to reduce the maximum beam size in the far field and broadens the minimum. It leads to a maximum beam diameter that decreases with the harmonic order in opposite to our experimental observations on harmonics 29 to 37.

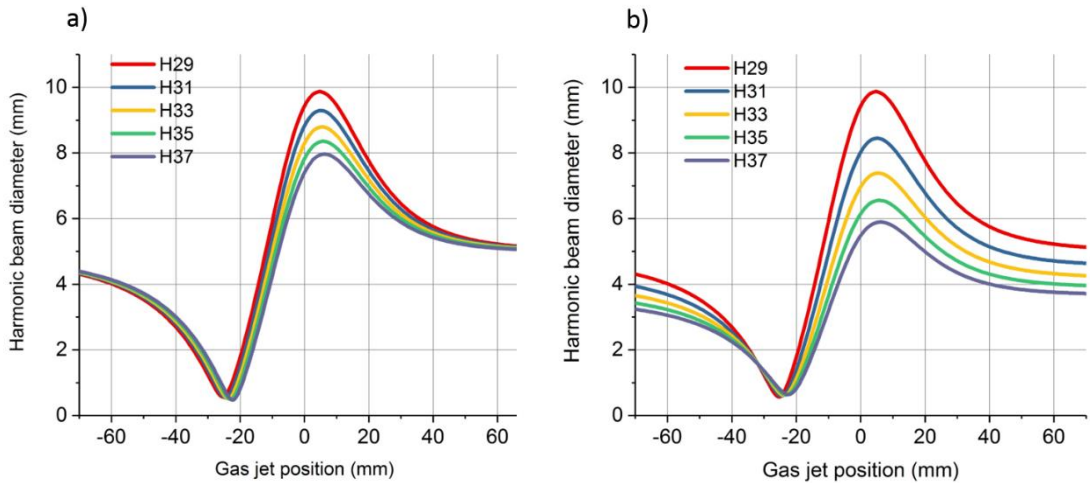

**Fig. S6: Parametric evolution of the XUV beam size.** (a) influence of the harmonic wavelength when the alpha parameter is constant ($\alpha = 5$, $q_{eff} = 4.7$ for all harmonics), and (b) influence of the parameter $q_{eff}$ on the simulated beam size ($q_{eff}$ is 4.7 for H29 and increases by 1 between each harmonic, $\alpha = 5$ for all harmonics).

## S6 Impact of the longitudinal evolution of $\alpha$.

When we included the fact that $\alpha$ depends also on the IR intensity (Fig. S7 a) and therefore changes with z, the maxima and minima were shifted. We observed that a second extremum could appear at the place where the harmonic gets in the cutoff and where $\alpha$ reaches its asymptotic value (Fig. S7 b). For z < 0 it leads to the appearance of a second minimum and for z > 0 it leads to a shift of the position where the beam has its maximum size. We observed that this behavior depends on the specific evolution of $\alpha$ with intensity but the global trend was preserved and showed a far field beam size having one (or two) minimum when the jet is before the focus and one maxima when the jet is after the IR focus.

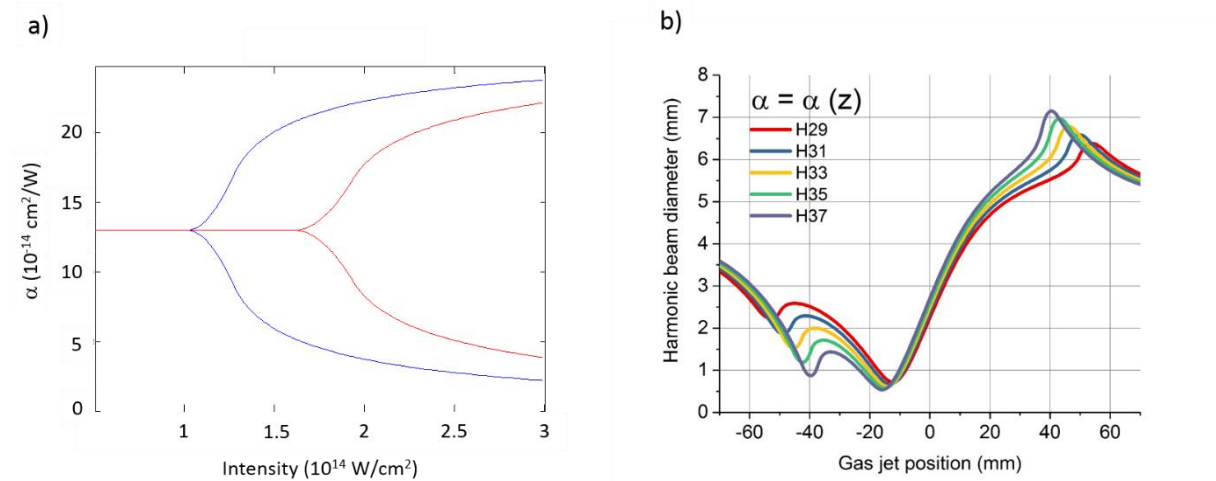

**Fig. S7: Influence of the $\alpha$ parameter.** (a) Intensity dependence of the $\alpha$ parameter used for this simulations for harmonic 29 (blue curve) and harmonic 37 (red curve). Only the lowest part of the curves corresponding to the short quantum path emission is considered in the simulations. (b) Simulated evolution of the diameters of the harmonic beams in the far field when considering that alpha changes with *z*.

This diameter evolution is also impacted by the peak intensity as shown in Fig. S8 for harmonics 29 and 37. We observe that changing the peak laser intensity at focus can affect the position at which the minimum beam size is observed. For harmonic 29, this position remains roughly constant since it is far from the position where this harmonic enters in the cutoff region (i.e. the two minima are well separated) and the minimum should remain well defined even after averaging over several intensities. For the highest harmonics (here 37) we observe that the two minima are closer and they can overlap at low intensity. When the position of the minima shifts with the intensity, averaging over several intensities should broaden the curve near its minimum or even make the minima disappear completely as observed experimentally with the highest harmonics.

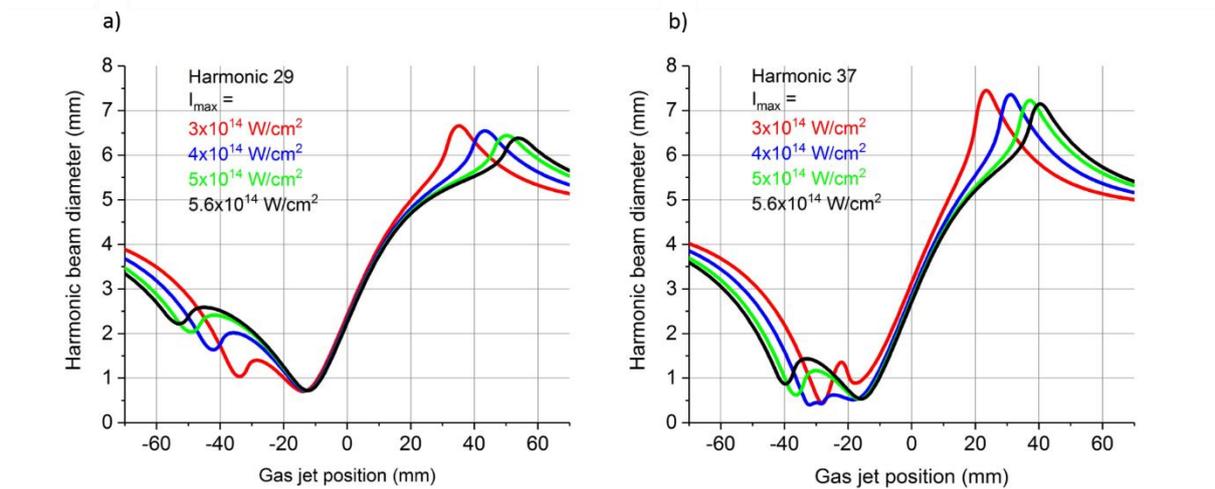

**Fig. S8: Beam size evolution for several peak intensities.** Evolution of the far field diameters of the 29 (a) and 37 (b) harmonic beams simulated with $\alpha$ changing with the longitudinal position for several peak laser intensities at focus.

**S7 Normalised foci shift.**

It is shown that the focus shift of consecutive harmonics is important for all position of the gas jet and we observe experimentally that the foci shift can be larger than the XUV confocal parameter when focusing is achieved. Our simulations show that the foci shift is important for all jet position. This is illustrated in Fig. S9 where we plot the difference in foci position between two consecutive harmonics (order q and q+2) divided by the Rayleigh length of the harmonic with order q. For this simulation, we consider that $\alpha$ evolves like in the main paper (linearly from $5 \times 10^{-14}$ cm$^2$/W for harmonic 29 to a constant value of $13 \times 10^{-14}$ cm$^2$/W for harmonic 39 and above).

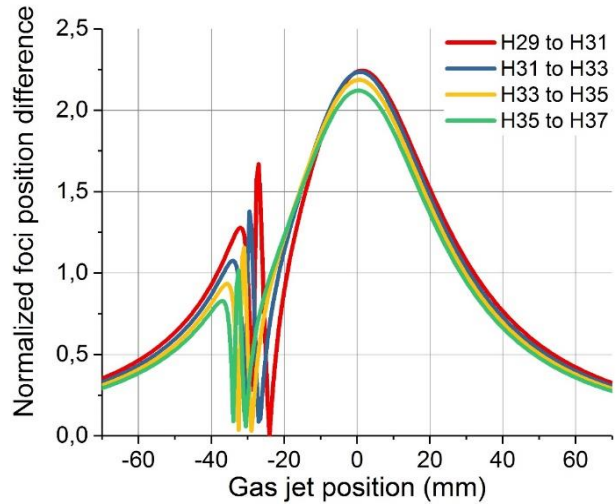

**Fig. S9: Normalised foci shift.** Value of the harmonic foci shift between two consecutive harmonics divided by the harmonic Rayleigh range. Here alpha changes with the harmonic order but does not depend on z.

We observe that the normalized relative shift is larger than 2 at z = 0 and remains significant for all jet positions. Around $z_{jet}$ = - 30, it is possible to have a zero relative shift between two specific harmonics but the shift is close to $Z_r$ for the next harmonics. This shift is related to $\alpha$ I and we observe that the shift is the strongest around z=0 where I is maximum. When we consider that alpha also increases at large z when a specific harmonic is near the cutoff, the foci shift further increases.

The Gaussian model was developed to highlight the physics that is behind the evolution of the XUV beam evolution. It is based on several hypotheses that help the understanding and simplify the generation process. It could nevertheless be interesting to have a more complete description that includes the spatio-temporal evolution of the IR intensity and the corresponding evolution of the harmonic phase and profile in the emission plane. Complete simulations could also include propagation in the generating medium even if propagation often modulates the efficiency without changing the XUV wavefront characteristics. This analysis goes beyond the scope of this paper but we performed model simulations with a more complete model dipole (where $\alpha$ changes with *r* and *t*) and with SFA model simulating HHG in a thin medium. The results obtained with two formalisms are shown in the following and exhibit the same trends as the one presented in this paper.

## S8 Simulations with non-Gaussian XUV beams.

In the main text of the document, we have developed a model assuming that the spatial profiles of the XUV beam are Gaussian. We have omitted, on purpose, the time evolution of the IR beam and obtained analytical formula.

In order to further test the results obtained with this model, we have performed more sophisticated simulations including the time evolution of the IR beam and also the XUV diffraction after the medium.

Two sets of simulations have been developed first with a model dipole having both $\alpha$ and $q_{eff}$ that evolve with time and space and second with the SFA model.

In this first study, the model dipole is written:

$$d(r,t) = I(r,t)^{q_{eff}(I(r,t))} e^{i\alpha(I(r,t))I(r,t)} \quad \text{Eq. (1)}$$

where *I(r,t)* is the time and space evolution of the IR beam. We follow the procedure described in (22) to calculate the far field spatially resolved harmonic spectra considering values of $\alpha$ (I(r,t)) for short trajectories. In Fig. S10, we present the results describing the far field evolution for H29 as a function of the gas jet position.

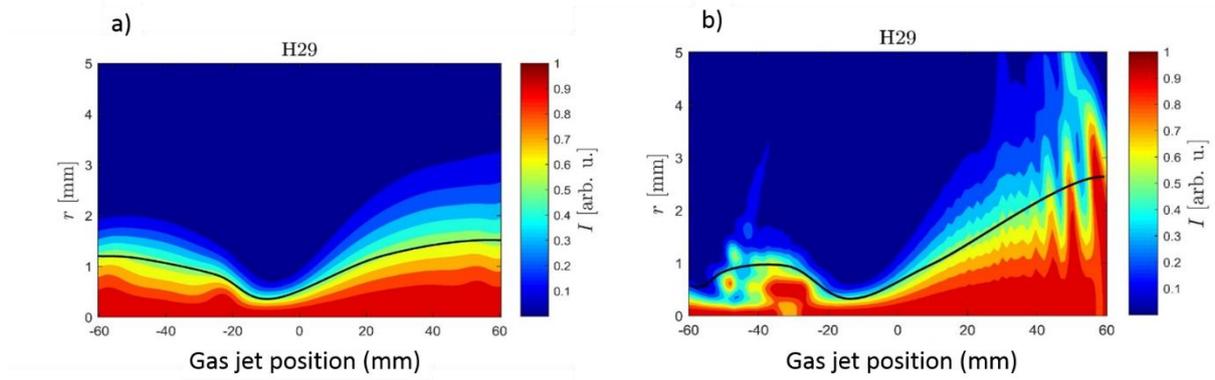

**Fig. S10: Non-Gaussian XUV beam size evolution.** a) Far field spatial distribution at the central frequency of H29 (integrated in frequency over 20 % of the harmonic width) for the dipole-model (short trajectory only). b) same as a) but for the SFA model without resolving the short-long trajectories. Both calculations have been performed in neon and at a peak intensity of $5 \times 10^{14}$ W/cm². The harmonic H29 has been considered for the simulation and the $q_{eff}$ and the $\alpha$ values as defined by Eq. (1) have been calculated from SFA using the saddle point approximation. Note that in this second dipole-model the $q_{eff}$ and the $\alpha$ are not constant value. The full black line is the FWHM of the far field-results averaged over intensity. The peak intensity distribution has been modeled by a Gaussian distribution which mimics the laser fluctuations.

We observe that the maximum size of the beam reaches smaller values in these simulations but the overall trend observed experimentally and detailed with the Gaussian model are qualitatively reproduced. In particular, a minimum of the beam size appear (here at -10 mm) when the jet is before the IR focus and it is followed by a maximum when the jet is after the focus (here at 60 mm). The spatial distribution appears more wiggled for the SFA model because of interferences appearing in the spectrum as described in (22). After averaging out the distribution over intensity, we obtain the full curve corresponding to the FWHM of the far field spatial distribution that shows the typical trend presented in the paper. Overall, the same trend has been obtained with several models (Gaussian, saddle point with effective value of $\alpha$ and $q_{eff}$, SFA) showing the robustness of the physics that defines the harmonic beam properties as described in this paper.